\documentclass[aps,pra,twocolumn, 10pt, showpacs,superscriptaddress,groupedaddress]{revtex4-1}  

\newcommand{\ie}{\emph{i.e.} }

\newcommand{\D}{\mathcal{D}}

\newcommand{\der}[2]{\ensuremath{\frac{d #1}{d #2}}}

\newcommand{\Rb}{\ensuremath{^{87}}Rb }
\newcommand{\ket}[1]{\ensuremath{\left| #1 \right>}}

\newcommand{\braket}[1]{\ensuremath{\left< #1 \right>}}
\newcommand{\eq}[2][{}]{\begin{equation}\label{#1} #2 \end{equation}}

\usepackage{graphicx}
\usepackage{epstopdf}
\usepackage{bbold}
\usepackage{comment}
\usepackage{color}
\usepackage{soul}
\usepackage[abs]{overpic}
\usepackage{amsmath,amsfonts,amssymb}
\usepackage{lipsum}
\usepackage{hyperref}
\usepackage{setspace}
\graphicspath{{Figures/}}
\begin{document}


%
\author{Gadi Afek}
\affiliation{Department of Physics of Complex Systems, Weizmann Institute of Science, Rehovot 76100, Israel }
\author{Jonathan Coslovsky}
\affiliation{Department of Physics of Complex Systems, Weizmann Institute of Science, Rehovot 76100, Israel }
\author{Arnaud Courvoisier}
\affiliation{Department of Physics of Complex Systems, Weizmann Institute of Science, Rehovot 76100, Israel }
\author{Oz Livneh}
\affiliation{Department of Physics of Complex Systems, Weizmann Institute of Science, Rehovot 76100, Israel }
\author{Nir Davidson}
\affiliation{Department of Physics of Complex Systems, Weizmann Institute of Science, Rehovot 76100, Israel }

\title[]{Observing Power-Law Dynamics of Position-Velocity Correlation in Anomalous Diffusion}


\begin{abstract}

In this letter we present a measurement of the phase-space density distribution (PSDD) of ultra-cold \Rb atoms performing 1D anomalous diffusion. The PSDD is imaged using a direct tomographic method based on Raman velocity selection. It reveals that the position-velocity correlation function $C_{xv}(t)$ builds up on a timescale related to the initial conditions of the ensemble and then decays asymptotically as a power-law. We show that the decay follows a simple scaling theory involving the power-law asymptotic dynamics of position and velocity. The generality of this scaling theory is confirmed using Monte-Carlo simulations of two distinct models of anomalous diffusion.

\end{abstract}


\maketitle


The phase-space density distribution (PSDD) contains information concerning the degrees of freedom of a system and allows calculation of any observable. An intriguing system to look at in this context is that of anomalous dynamics for which the mean square displacement (MSD) scales as $\braket{x^2}\sim t^{2\alpha}$, with $\alpha\neq 1/2$. This type of dynamics, found in a wide variety of systems in nature ranging from dynamics of ``bubbles" in denaturing DNA molecules~\cite{Bar2007}, through fluctuations in the stock-market~\cite{Plerou2000} to models describing brief awakenings in the course of a night's sleep~\cite{Lo2002}, is generally non-universal and system-dependent~\cite{Metzler2000,Sokolov2012,Zaburdaev2015}. 

A uniquely interesting model system for the study of anomalous diffusion is that of cold atoms diffusing in a dissipative 1D lattice, closely related to L\'evy walks and motion in logarithmic potentials, displaying such phenomena as the breakdown of ergodicity and of equipartition, memory effects and slow relaxation to equilibrium~\cite{Castin1991,Marksteiner1996,Katori1997,Lutz2003,Lutz2004,Jersblad2004,Douglas2006,Kessler2010,Hirschberg2011,Sagi2012,Kessler2012,Dechant2012,Barkai2014,Dechant2014,Zaburdaev2015,Holz2015,Dechant2015PRL,Dechant2016}. The major advantage of such a system is the high degree of control it enables over the physical parameters governing the dynamics. One of the fundamental insights that can be obtained from the PSDD of such a system is the phase-space cross correlation between position and velocity $C_{xv}$. $C_{xv}$ can reveal the fingerprint of the underlying model and in particular is essential for understanding concepts and techniques such as adiabatic cooling in lattices~\cite{Kerman2000}, stochastic cooling~\cite{Raizen1998}, point source atom interferometry~\cite{Dickerson2013,Hoth2016} and enhanced velocity resolution~\cite{Henson2012,Damon2014}, alongside elementary notions in quantum mechanics~\cite{Robinett2005}. These correlations have been surprisingly overlooked in both theory and experiment, perhaps due to the lack of a direct method for imaging the phase-space of atomic clouds, which does not require cumbersome mathematical tools or a specific potential~\cite{Kurtsiefer1997,Pfau1997,DelCampo2008,Lee2012,Zhou2014}. No analysis of the dynamics of the correlations has been reported to the best of our knowledge.

In this letter we analyze and measure the dynamics of the position-velocity correlation of an ensemble of classical particles, originating from a point-like source and undergoing one-dimensional anomalous super-diffusion. The measurement is done using a new tomographic method for direct phase-space imaging, utilizing a combination of the straightforward tools of absorption imaging and the velocity-sensitivity of Raman control. We obtain qualitative agreement with theory, in the form of a scaling argument we derive, connecting the temporal asymptotics of the correlations with those of position and velocity. We verify the universality of the scaling-law using two different types of Monte-Carlo simulations.

The position-velocity correlation function is defined as-
\eq[eq:correlation_definition]{C_{xv}(t)\equiv\frac{\braket{\delta\vec{x}(t)\cdot\delta\vec{v}(t)}}{\sqrt{\braket{\delta\vec{x}^2(t)}\braket{\delta\vec{v}^2(t)}}},}
where for any observable $\mathcal{A}$ and PSDD $f(x,v)$, $\braket{\mathcal{A}} = \int \mathcal{A}f(x,v)dxdv$ and $\delta \mathcal{A} = \mathcal{A} -\braket{\mathcal{A}}$. Calculation for an initially uncorrelated ensemble of particles, reveals that $C_{xv}(t)$ asymptotically approaches unity for ballistic motion and decays as $1/\sqrt{t}$ for normal diffusion~\cite{Gillespie2012}. The inherent timescales depend strongly on the initial conditions of the ensemble, and their observation demands a point-like atomic source.

For the general case of power-law dynamics and anomalous diffusion, we use eq.~\ref{eq:correlation_definition} to derive a scaling argument, assuming power-law behaviour of both $\braket{\delta x^2(t)}\sim t^{2\alpha}$ and $\braket{\delta v^2(t)}\sim t^{2\beta}$. The scaling of the numerator of eq.~\ref{eq:correlation_definition} is calculated by taking the derivative
\eq[eq:derivative]{\braket{\delta x\cdot \delta v} \sim \der{\braket{\delta x^2}}{t} \sim t^{2\alpha-1}}
The denominator of eq.~\ref{eq:correlation_definition} gives $t^{\alpha+\beta}$. Together this yields 
\eq[eq:scaling]{C_{xv}(t)\sim t^{\alpha-\beta-1}\sim t^\gamma.}
The $(\alpha,\beta) = (1/2,0)$ normal-diffusive limit and $(\alpha,\beta) = (1,0)$ ballistic limit give $\gamma=-1/2$ and $\gamma=0$ respectively, indicating a decay of $1/\sqrt{t}$ for normal-diffusion and saturation at a nonzero value for ballistic motion. 

\begin{figure}
	\centering
	\begin{overpic}
		[width=\linewidth]{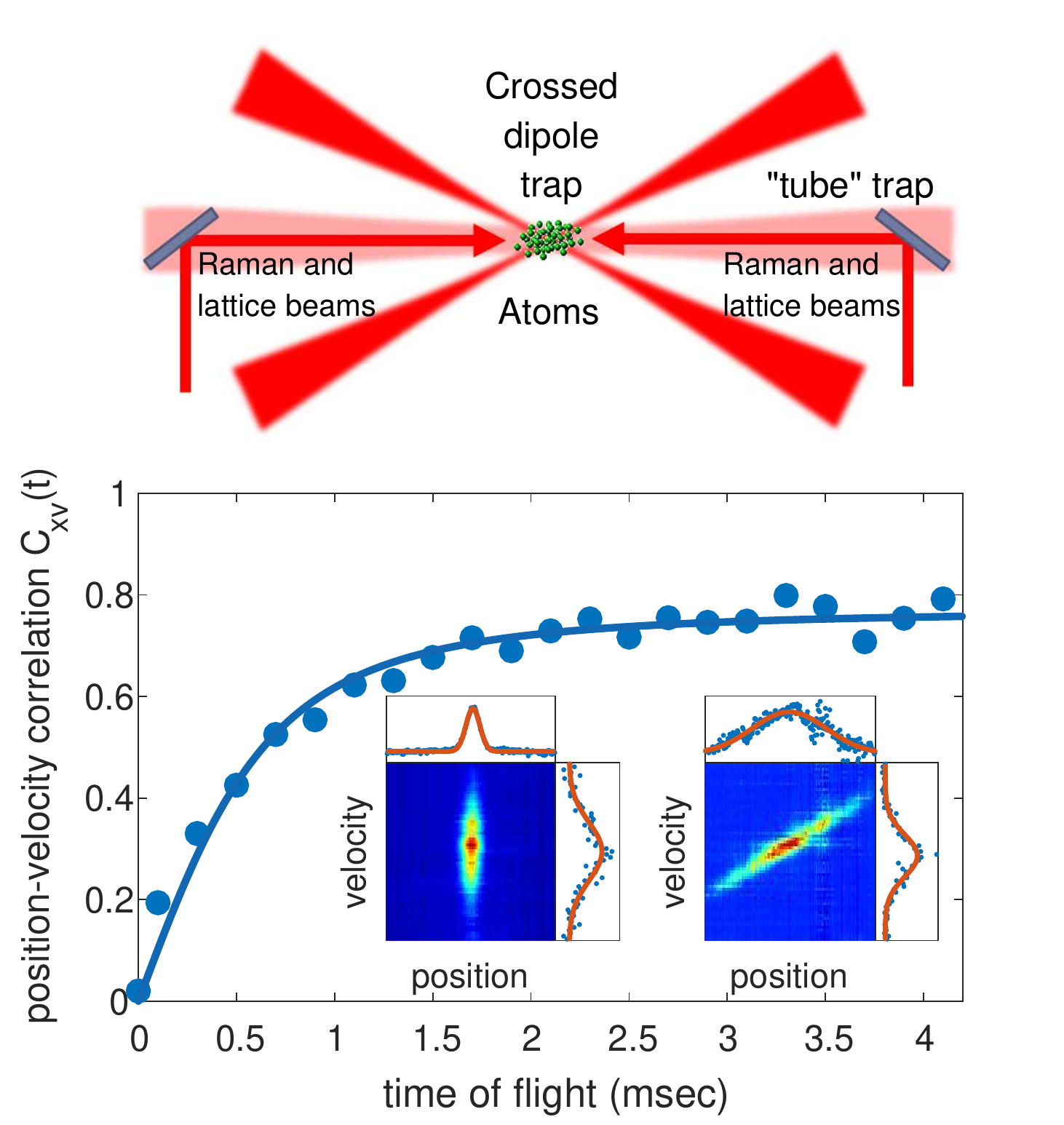}
		\put(35,240){\large \textbf{(a)}}
		\put(35,130){\large \textbf{(b)}}
	\end{overpic}
	\caption{\textbf{(a)} A sketch of the experimental setup. Laser-cooled \Rb atoms are loaded into a crossed optical dipole-trap and evaporated. They are then transferred into a single-beam red-detuned tube trap and perform anomalous diffusion in a dissipative optical lattice at a certain lattice depth for a given time. Their phase-space distribution is then measured using the tomographic method described in the text. \textbf{(b)} Measured position-velocity correlation for ballistically expanding atoms. Solid line represents the fit to eq.~\ref{eq:ballistic_correlation_omega}. Left (right) insets show short (long) time phase-space reconstruction. Integrals over the velocity and position axes are presented along with a fit to a Gaussian, taken for ballistic expansion after 0.1 (4.1)~msec time-of-flight. The shearing of phase-space indicates correlations between position and velocity.}
	\label{fig:fig1}
\end{figure}

In the experiment, (fig.~\ref{fig:fig1} (a)) a cloud of $\sim10^5$ $^{87}$Rb atoms is loaded into a crossed dipole trap from a Raman-sideband cooled magneto-optical trap. After a short evaporation and thermal equilibration stage the point-like atomic cloud ($\sim30\mu$m in size, $T\approx 10\mu$K) is loaded adiabatically into a single-beam, elongated dipole trap providing confinement in the radial axis (for details see~\cite{Sagi2012}). The atoms then undergo anomalous superdiffusion for lattice exposure time $t$, in a 1D Sisyphus lattice of depth $U_0$~\footnote{It has been shown in~\cite{Castin1991} that the dipole potential depth of the lattice is the sole parameter governing the asymptotic dynamics}, originating from a distributed feedback diode laser (DFB) detuned -66 MHz relative to the transition between states $5^2S_{1/2}$, $F = 2$ and $5^2P_{3/2}$, $F'=3$.

We then perform tomographic phase space imaging by transferring atoms whose velocity lies within a narrow velocity class, from the \ket{F=1} lower hyperfine ground state to the upper ground-state level \ket{F=2} using a Raman velocity-selective $\pi$-pulse with two counter-propagating beams. The center of the selected velocity class is scanned by varying the the two-photon detuning of the pulse, and the Rabi frequency sets its width~\cite{Kasevitch1991}. These atoms are imaged onto a CCD camera using state-selective absorption imaging. The measured PSDD is depicted, for ballistic expansion (\ie with $U_0=0$), in the left and right insets of figure~\ref{fig:fig1} (b) for short (0.1~msec) and long (4.1~msec) times respectively, revealing the expected shearing of the PSDD. The position-velocity correlation is extracted from the data~\footnote{See supplementary material} and shown in figure~\ref{fig:fig1} (b) as a function of free propagation time $t$ (termed ``time of flight" for $U_0=0$). 
Correlations brought about by ballistic expansion, starting with an uncorrelated Gaussian phase-space are given by~\cite{Robinett2005}
\eq[eq:ballistic_correlation_omega]{C_{x,v}(t) = \frac{\omega_{\text{osc}}t}{\sqrt{1+\omega_{\text{osc}}^2t^2}}}
where $\omega_{\text{osc}}\equiv \sigma_v(t=0)/\sigma_x(t=0)$ sets, under thermal equilibrium, the ratio between the initial standard-deviation of the velocity distribution, $\sigma_v$ and that of the position distribution $\sigma_x$. It also represents the oscillation frequency in the trap prior to the release. This parameter sets the initial slope of the build-up of the correlations.

To establish the validity of the new tomographic method, we test it on this textbook case. We fit the ballistic expansion phase-space tomography shown in fig~\ref{fig:fig1} (b) to eq.~\ref{eq:ballistic_correlation_omega}, and obtain a value of $\omega_{\text{osc}}=2\pi\times (213\pm15)$~Hz, in excellent agreement with the value obtained independently of $\omega_{\text{osc}}=2\pi\times (230\pm3)$~Hz, measured by giving a small kick to the trapped atoms, and imaging the oscillations in the trap. The measured zero-time correlation value is consistent with zero up to the measurement error, as expected from an equilibrated cloud. The saturation value obtained from the fit, $C_\infty = 0.77\pm0.01$, is sub-unity due to a broadening effect arising from a finite two-photon Rabi frequency required to obtain good SNR. The broadening in the correlation is a function of the ratio between the spectral width of the velocity-selective Raman $\pi$-pulse (rescaled by $2k_L$, the wavenumber of the Raman laser) and that of the velocity distribution. As the Rabi frequency becomes small compared to the width of the velocity distribution the measured correlation becomes closer to the real value. The Rabi frequency selected for the experiment reflects the tradeoff between minimizing the broadening and obtaining good SNR. Calibrating this effect we rescale the data such that $C_\infty = 1$.
\begin{figure}
	\centering
	\begin{overpic}
		[width=\linewidth]{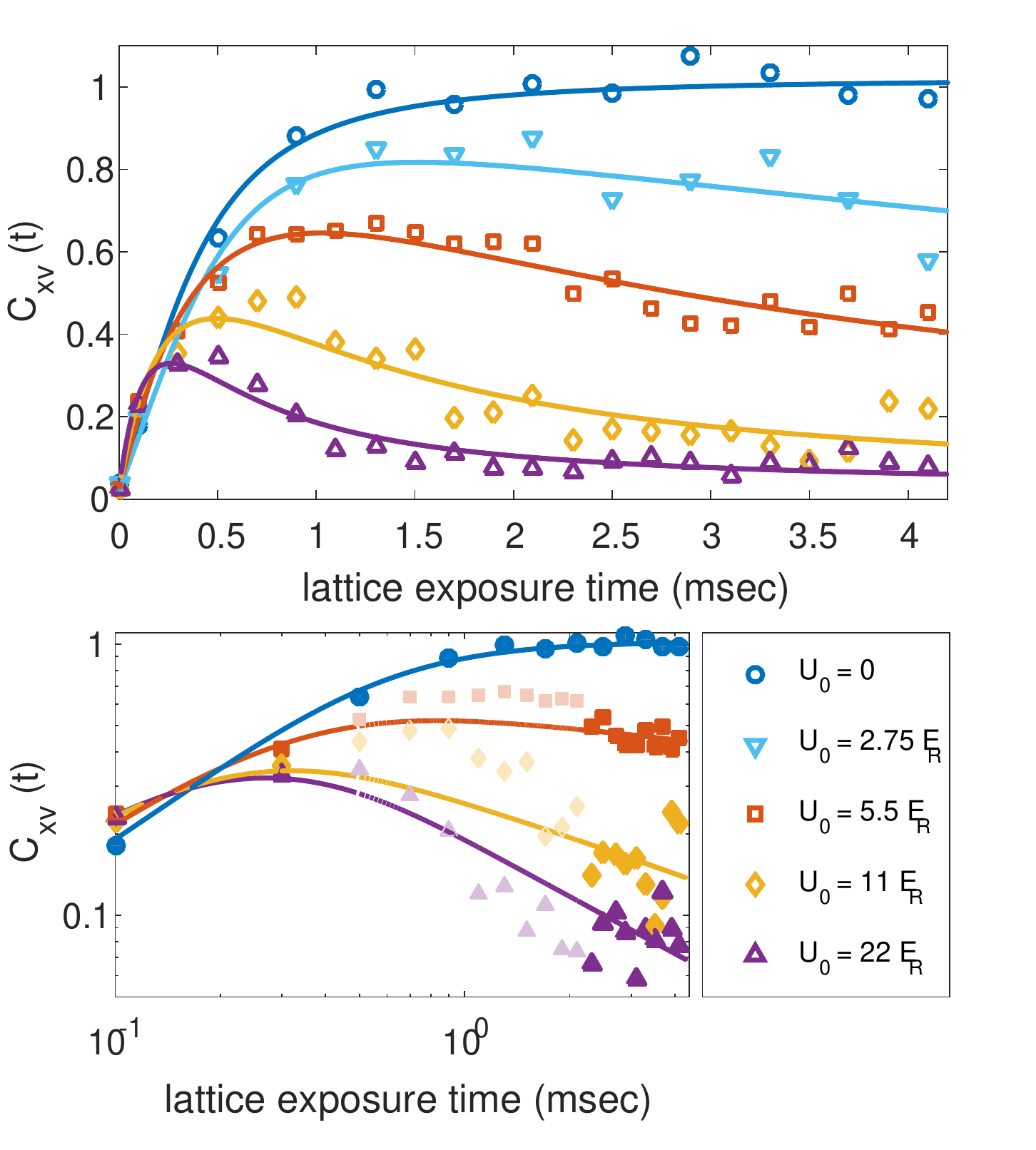}
		\put(30,245){\large \textbf{(a)}}
		\put(30,110){\large \textbf{(b)}}
	\end{overpic}
	\caption{\textbf{(a)} Rescaled position-velocity correlations as a function of lattice exposure time and lattice depth. Colors and symbols indicate different lattice depths, $U_0$. $E_\text{R}$ is the recoil energy. At short times the correlations build up and are later quenched at varying rates, depending on the anomalous dynamics. Solid lines indicate the fit to the interpolation formula of eq.~\ref{eq:main_gen}. \textbf{(b)} Fit to the interpolation formula of eq.~\ref{eq:main_gen} on a log-log scale, excluding interim times $0.5\leq t\leq2.1$~msec. Shaded symbols are excluded points. The $U_0=0$ data set has no exclusions.}
	\label{fig:fig2}
\end{figure} 
Figure~\ref{fig:fig2} presents the measured, rescaled position-velocity correlations as a function of lattice exposure time $t$ and lattice depth in linear (a) and log-log (b) scales, ranging from ballistic to normal diffusion. It shows the initial build-up and sequential decay of the correlations. 

To develop a theoretical description, we first consider the limits of normal diffusion and ballistic expansion, using the Langevin equation approach to normal Brownian motion~\cite{Pathria1996,Luczka2000,Gillespie2012}. The instantaneous acceleration of a particle in a medium is given by $\dot{\vec{v}} = -\Gamma\vec{v} + \vec{\mathcal{A}}$, where $\vec{v}$ is the velocity vector, $\Gamma$ is the drag coefficient setting the timescale for transition between the ballistic and diffusive regimes and $\vec{\mathcal{A}}$ is the Langevin random acceleration. For simplicity we assume $\braket{x}=\braket{v}=0$, hence $\delta x = x$ and $\delta v = v$.
The numerator of eq.~\ref{eq:correlation_definition} can be calculated by noticing that-
\begin{equation}
\label{eq:diff}
		\der{}{t}\braket{\vec{x}\cdot\vec{v}} = -\Gamma\braket{\vec{x}\cdot\vec{v}}+\braket{\vec{v}^2}
\end{equation}
where due to the randomness of the Langevin acceleration $\braket{\vec{x}\cdot\vec{\mathcal{A}}}=0$. $\braket{\vec{v}^2}$ is given by-
\eq[eq:vsq]{\braket{\vec{v}^2} = \sigma_{v_0}^2 + (\sigma_{v_\text{eq}}^2-\sigma_{v_0}^2)(1-e^{-2\Gamma t})}
with $\sigma_{v_0}^2$ denoting the initial variance of the velocity distribution, and $\sigma_{v_\text{eq}}^2 = \sigma_v^2(t\to\infty)$. Substituting into eq.~\ref{eq:diff} and solving under an uncorrelated initial condition yields
\eq[eq:xv]{\braket{\vec{x}(t)\cdot\vec{v}(t)} = \frac{e^{-2\Gamma t} \left(e^{\Gamma t}-1\right)\left[\sigma_{v_\text{eq}}^2\left(e^{\Gamma t}-1\right)+\sigma_{v_0}^2\right]}{\Gamma}.}
Calculating the terms in the denominator~\cite{Pathria1996,Gillespie2012} under the initial conditions $\braket{\vec{x}^2(0)} = \sigma_{x_0}^2$ and $\der{\braket{\vec{x}^2}}{t}\vert_{t=0}=0$, and setting $\sigma_{v_0}^2 = \sigma_{v_\text{eq}}^2$ (see~\cite{Note2} for full expression) we obtain,
\begin{equation}
\label{eq:main}
	C_{xv}(t) = \frac{e^{-\Gamma t/2}\left(e^{\Gamma t}-1\right)}{\left[2+e^{\Gamma t}\left((\Gamma/\omega_{\text{osc}})^2+2\Gamma t-2\right)\right]^{1/2}}.
\end{equation}
Equation~\ref{eq:main} reveals the initial linear rise in correlation due to the ballistic timescale of the dynamics and the asymptotic decay $\sim t^{-1/2}$ of the normal-diffusive correlations. The ballistic regime (eq.~\ref{eq:ballistic_correlation_omega}) is obtained from it by taking the $\Gamma\to0$ limit. 

Equation~\ref{eq:main} can be generalized to account for anomalous diffusion and hence the power-law decay anticipated by eq.~\ref{eq:scaling} as-
\begin{equation}
\label{eq:main_gen}
	C_{xv}(t) = \frac{e^{-\Gamma t/2}\left(e^{\Gamma t}-1\right)}{\left[2+e^{-\frac{\Gamma t}{2\gamma}}\left[\left(\Gamma/\omega_{\text{osc}}\right)^{-1/\gamma}+2\Gamma t-2\right]\right]^{-\gamma}}.
\end{equation}
This preserves $C_{xv}\sim t^\gamma$ at long times and the initial $C_{xv}\sim t$ at short times. It recovers eq.~\ref{eq:main} for $\gamma=-1/2$. In figure~\ref{fig:fig2} (a,b) we show in solid lines the fit of this function to the data with $\gamma$, $\omega_{\text{osc}}$ and $\Gamma$ as fit parameters~\cite{Note2}. There exist two time \emph{scales} and two temporal \emph{scalings}. The buildup scales linearly in time and saturates at unity with a timescale of $1/\omega_{\text{osc}}$. The decay scales like $t^\gamma$ with a timescale of $1/\Gamma$. The transition between the buildup and decay occurs at a timescale $\tau_m$, which is approximately the average $\left(1/\omega_{\text{osc}}+1/\Gamma\right)/2$~\cite{Note2}. It is henceforth evident that observing the short-time correlation dynamics requires $\tau_m$ to be within the measurement time.  

\begin{figure}
	\centering
	\begin{overpic}
		[width=\linewidth]{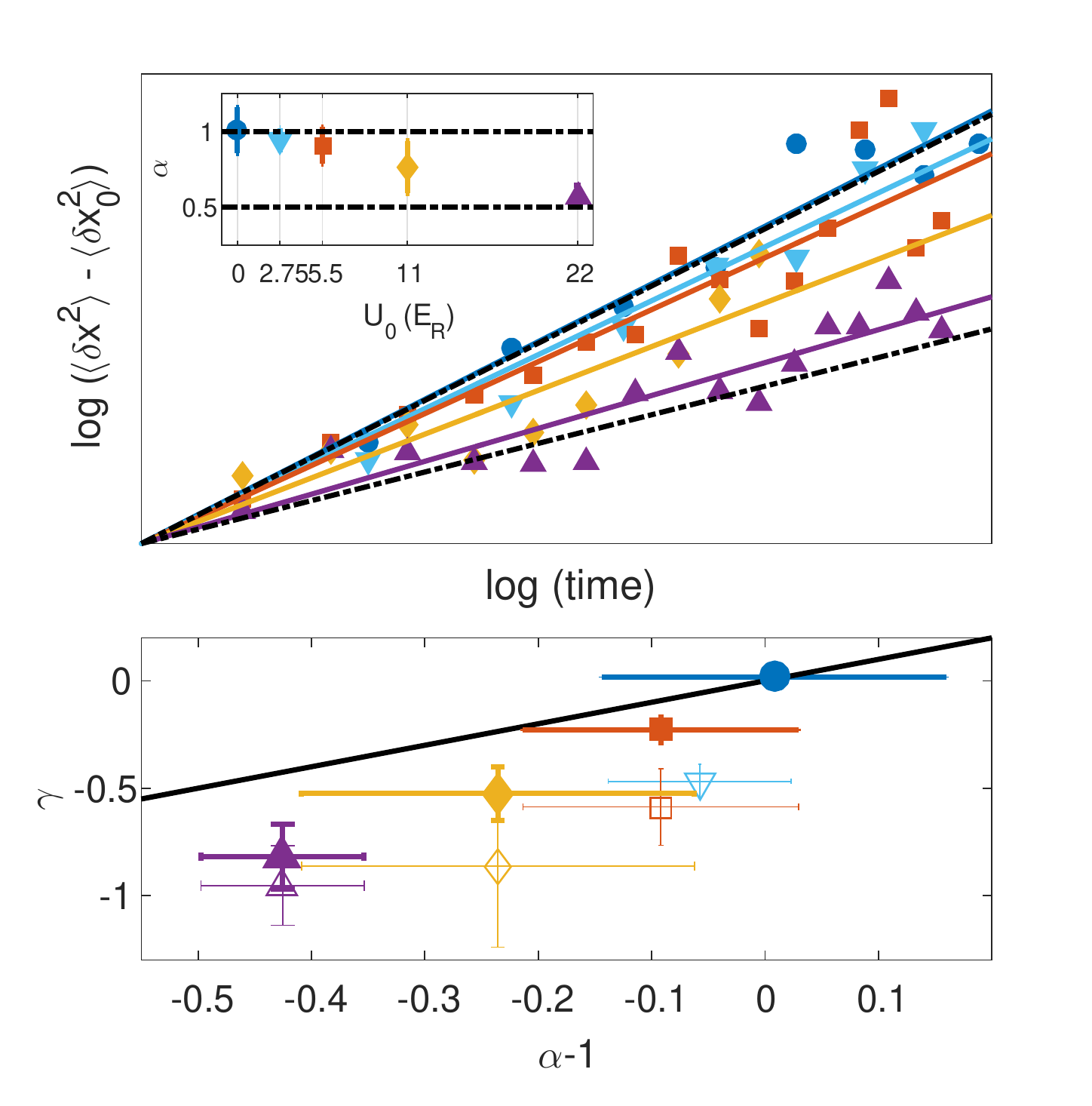}
		\put(200,140){\large \textbf{(a)}}
		\put(200,40){\large \textbf{(b)}}
	\end{overpic}
	\caption{\textbf{(a)} The variance of the position as a function of time on a log-log scale for various lattice depths, $U_0$. Solid lines are linear fits, whose slope is summarized in the inset, bearing good qualitative agreement with~\cite{Sagi2012}. The data is shifted to cross at the origin. Black dotted lines represent the ballistic and normal-diffusive limits, showing that the entire superdiffusive regime is accessible. \textbf{(b)} Experimental demonstration of the scaling relation of eq.~\ref{eq:scaling} for the case of $\beta=0$ (relaxed velocity dynamics). The exponent of the correlation, $\gamma$, extracted from fitting the interpolation functionof eq.~\ref{eq:main_gen} to the data of fig.~\ref{fig:fig2} (a), is plotted in empty symbols as a function of $\alpha-1$. The full symbols represent a fit excluding $0.5\leq t\leq2.1$~msec. The colors and shapes correspond to the lattice depths as in fig.~\ref{fig:fig2}. The solid line is the theoretical scaling relation.}
	\label{fig:fig3}
\end{figure}

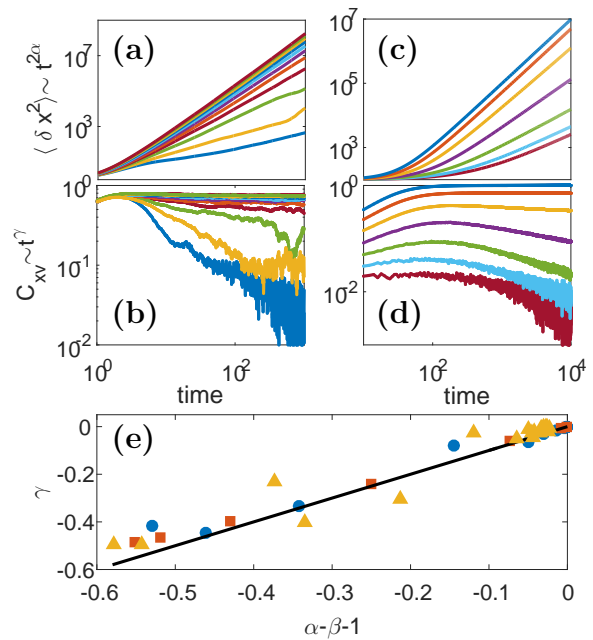
\begin{figure}
	\centering
	\begin{overpic}
		[width=\linewidth]{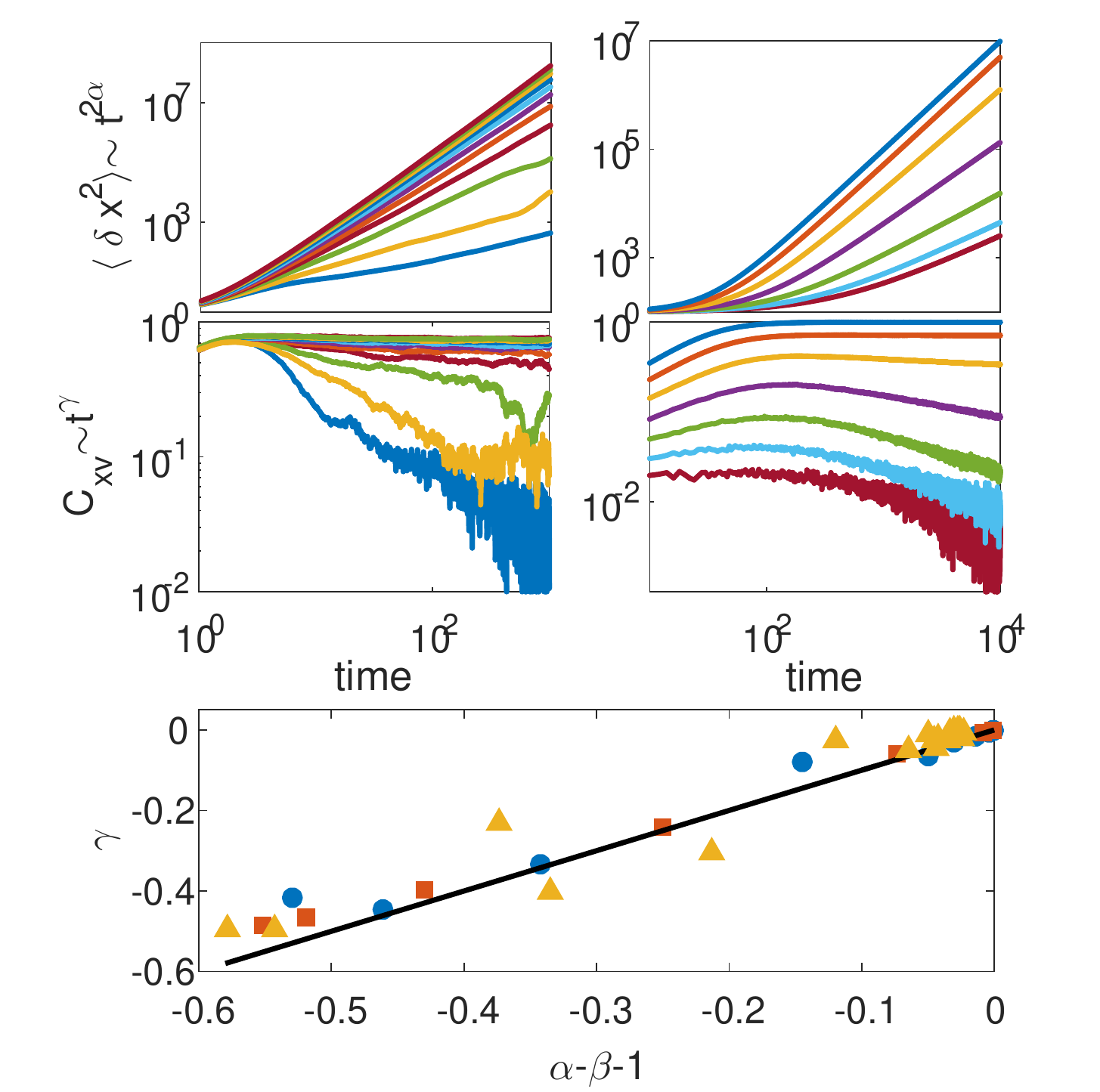}
		\put(50,220){\large \textbf{(a)}}
		\put(150,220){\large \textbf{(c)}}
		\put(50,120){\large \textbf{(b)}}
		\put(150,120){\large \textbf{(d)}}
		\put(50,73){\large \textbf{(e)}}
	\end{overpic}
	\caption{Results of numerical Monte-Carlo simulations of two anomalous diffusion models. \textbf{(a)} $\langle\delta x^2\rangle$ as a function of time for different lattice depths, corresponding to the range of superdiffusive behavior. Larger slopes correspond to shallower lattices. A linear fit to the long-time data enables extraction of $\alpha$. \textbf{(b)} Position - velocity correlations, showing the predicted power-law dependence. Stronger decay corresponds to deeper lattice. (a, b) are obtained using the Langevin simulation of eq.~\ref{eq:Langevin_discrete}. \textbf{(c, d)} Similar behavior, obtained from the L\'evy walk simulation. Larger slopes correspond to smaller $\gamma_0$ in \textbf{(c)}, stronger decay corresponds to larger $\gamma_0$ in \textbf{(d)}. \textbf{(e)} Validation of the scaling relation $\gamma=\alpha-\beta-1$. Light-blue circles: pre-equilibrated dynamics in the Langevin simulation ($\beta=0$), red squares: L\'evy simulation (inherently pre-equilibrated, $\beta=0$) and orange triangles: velocity dynamics in the Langevin simulation ($\beta\neq0$).}
	\label{fig:fig4}
\end{figure}

Figure~\ref{fig:fig3} (a) presents the position variance $\braket{\delta x^2(t)}$, for various lattice depths. The position distribution is obtained by integrating over the velocity axis of the tomographic phase-space images (see insets of fig.~\ref{fig:fig1} (b)). Fitting $\braket{\delta x^2(t)}-\braket{\delta x^2(0)}\sim t^{2\alpha}$ reveals that the entire superdiffusive regime is accessible in the experiment, as seen in the inset, bearing qualitative agreement with~\cite{Sagi2012}. Fig.~\ref{fig:fig3} (b) presents the decay exponent $\gamma$ extracted from the fits of fig.~\ref{fig:fig2} as a function of $\alpha-1$. The velocity distribution equilibrates at a fast timescale ($1/\Gamma<1$~msec) to a steady-state value, meaning $\beta\approx0$~\footnote{Theory predicts $\beta>0$ for shallow lattices, however in our limited measurement-time we cannot observe this.}. The results (empty symbols) follow the trend of the scaling argument prediction of eq.~\ref{eq:scaling} but are significantly beneath it. Excluding the intermediate times $0.5\leq t\leq2.1$~msec from the fit yields qualitatively similar results, but with better agreement to the scaling argument (full symbols). This indicates that our interpolation function (eq.~\ref{eq:main_gen}) describes well the short and long time dynamics but fails to describe the intermediate times.
To test the generality of our scaling argument we numerically study the dynamics of the position-velocity correlations within the framework of two distinct models featuring anomalous diffusion. The first describes semiclassical atomic motion in a 1D Sisyphus lattice~\cite{Barkai2014}, using the Langevin phase-space equations
\begin{equation} \label{eq:Langevin_discrete}
        \dot x = v,\quad\dot v = -\frac{v}{1+v^2} + \sqrt{2\D}\xi(t)
\end{equation}
The white noise term $\xi(t)$ is Gaussian and has zero mean. The initial conditions are Gaussian, uncorrelated distributions of standard deviation $\sigma=1$ in both velocity and position. The diffusion constant is related to the depth of the lattice by~\cite{Castin1991,Marksteiner1996} $\D = c E_\text{R}/U_0$, where $c$ is a dimensionless parameter of order 10. Figure~\ref{fig:fig4} (a) presents the simulated size of the cloud as a function of time and diffusion constant for $N=1\times10^4$ atomic trajectories. The power-law dependence $\braket{x^2}\sim t^{2\alpha}$ is evident. The width of the velocity distribution scales like a power law in time~\cite{Kessler2010}. The position-velocity correlation is calculated using the definition (eq.~\ref{eq:correlation_definition}), and shown in fig.~\ref{fig:fig4} (b). We fit the long-time decay of the correlation to $t^{\gamma}$ and plot $\gamma$ as a function of $\alpha-\beta-1$ in fig.~\ref{fig:fig4} (e), for two distinct scenarios, one where the velocities are initialized in some arbitrary initial size (orange triangles) and one where they are initialized at their steady-state value corresponding to each lattice depth, setting $\beta=0$ (blue circles). 
The second simulation is a L\'evy walk simulation~\cite{Zaburdaev2015}, where particles are initialized in an uncorrelated Gaussian phase-space and proceed to perform walks of durations $\tau$, drawn from a unity-scaled Lomax distribution $\psi_{\gamma_0}(\tau) = \gamma_0/(1+\tau)^{1+\gamma_0}$. The width of the velocity distribution remains constant throughout the simulation ($\beta=0$). $1<\gamma_0<2$ gives access to the superdiffusive regime. Figure~\ref{fig:fig4} (c) shows the size of the cloud as a function of time and fig.~\ref{fig:fig4} (d) the power-law decay of the correlations. Fig.~\ref{fig:fig4} (e) shows the summary of the relation between the exponents $\alpha$, $\beta$ and $\gamma$ obtained using this method (red squares). All the simulation results, for the two distinct anomalous diffusion models, agree well with theory, indicating the generality of our scaling argument (eq.~\ref{eq:scaling}).
In summary, we present a measurement of the initial build-up and sequential decay of position-velocity correlations for a system of cold atoms performing anomalous superdiffusion. We find that the correlations decay asymptotically with a power-law exponent relating to the power-law exponents of the position variance and the velocity variance, in qualitative agreement with a simple scaling argument we derive. The universality of the scaling law is validated using Monte-Carlo simulations of two distinct models of anomalous diffusion. This universal relation between the long time decay of $C_{xv}$ and other exponents that are easier to measure can be used to infer $C_{xv}$ for systems where it cannot be measured directly. The position-velocity correlations are obtained using a new direct method to measure the phase-space density distribution that can be used to access different types of phenomena such as deviations from equipartition theorem~\cite{Dechant2015PRL,Dechant2016} for the nonequilibrium steady-state scenario of the discussed system with the addition of an underlying harmonic potential, and to probe phase-space correlations in systems of a quantum nature, described by a single wavefunction~\cite{Robinett2005, Schweigler2017}. The short-time dynamics in anomalous diffusion is model-dependent and non-trivially experimentally accessible~\cite{Kheifets2014}. Our work invites theoretical analysis of the correlation function as a fingerprint of the details of the underlying model.

\begin{acknowledgments}
The authors would like to thank Eli Barkai, Andreas Dechant and Erez Aghion for valuable theoretical input and Yoav Sagi, Hagai Edri and Noam Matzliah for discussions.
\end{acknowledgments}

\bibliographystyle{apsrev4-1}
\bibliography{Correlations}
\end{document}



\heading{Supplementary Material for Observing Power-Law Dynamics of Position-Velocity Correlation in Anomalous Diffusion}
\begin{center} Gadi Afek, Jonathan Coslovsky, Arnaud Courvoisier, Oz Livneh and Nir Davidson\end{center}
\begin{center} Department of Physics of Complex Systems, Weizmann Institute of Science, Rehovot 76100, Israel\end{center}


\subsection*{Phase-space tomography and experimental sequence}

\begin{figure}
	\centering
		\begin{subfigure}{\textwidth}
		  \centering
		  \includegraphics[width = \textwidth]{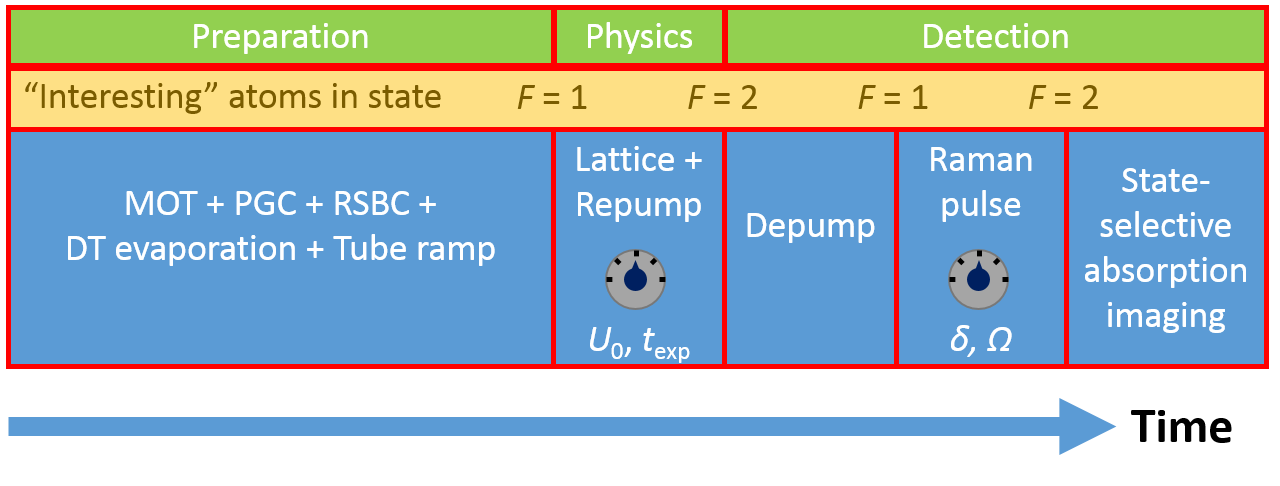}
		\end{subfigure}
	\caption[Scheme of the experimental sequence]{Scheme of the experimental sequence. $^{87}$Rb atoms are prepared in a crossed dipole trap and then loaded into a 1D "tube trap". Then a time-varied and depth-varied lattice pulse is given, followed be the detection stage comprised of a Raman velocity-selective pulse followed by absorption imaging.}
	\label{fig:sequence}
\end{figure}

\begin{figure}
	\centering
		\begin{subfigure}{.5\textwidth}
		  \centering
		\includegraphics[width = 60mm]{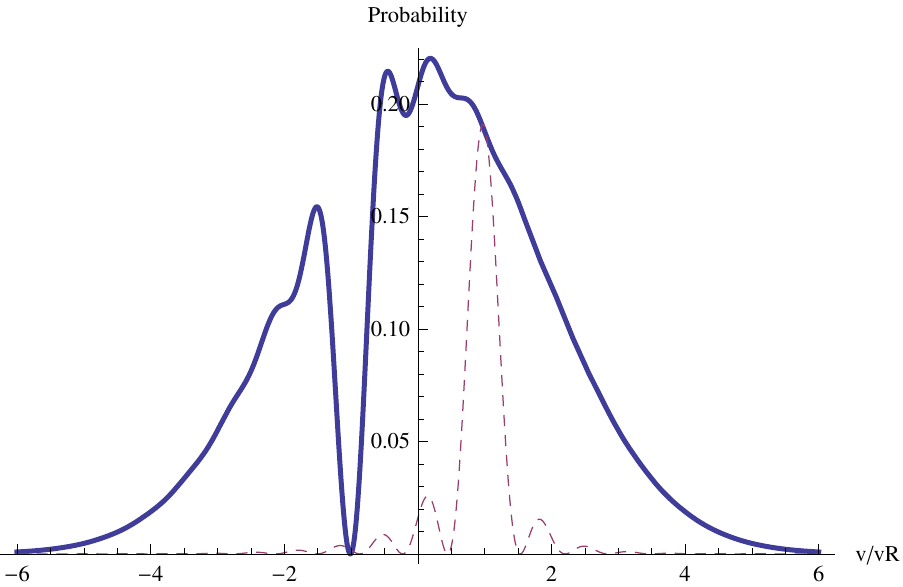}
		\end{subfigure}%
		\begin{subfigure}{.5\textwidth}
		  \centering
		  \includegraphics[width = 60mm]{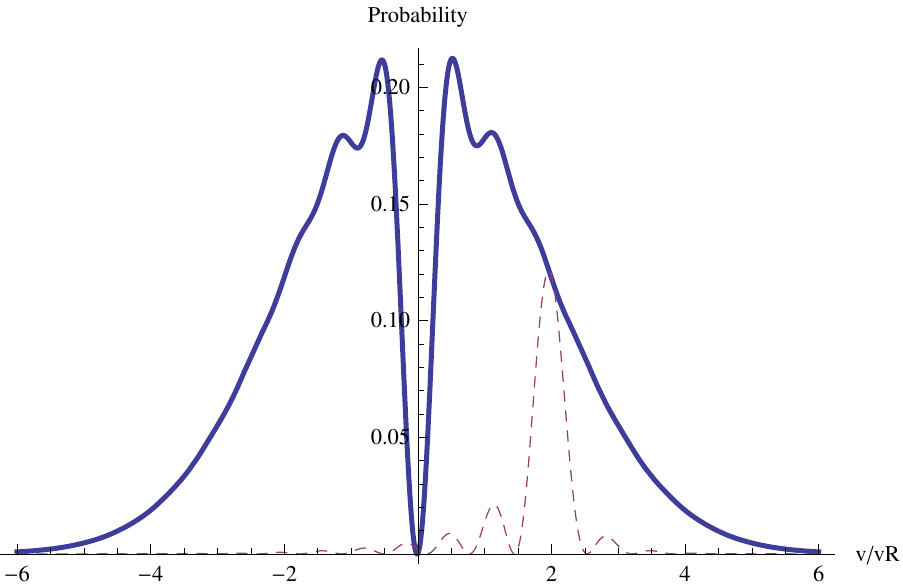}
		\end{subfigure}
	\caption[Analytic calculation of the distribution of momentum after a velocity selective $\pi$ pulse]{Analytic calculation of the distribution of momentum in \ket{1} and \ket{2} states (solid and dashed lines, respectively) after a $\pi$ pulse for two different "digging points" (left and right). Notice that a net momentum of twice the one-photon recoil momentum is imparted to the state \ket{2} in completing the hyperfine transition. The location of the "digging point" is a function of the detuning $\delta$. Detection of the transferred population can be deconvolved back to give the original (unknown) distribution}
	\label{Fig:Gaussians}
\end{figure}

The experimental sequence, depicted in figure~\ref{fig:sequence} is as follows: A cloud of $\sim10^5$ $^{87}$Rb atoms are loaded into a crossed dipole trap from a Raman-sideband cooled, polarization gradient cooled MOT. After a short evaporation and equilibration phase the atoms are loaded adiabatically into a single-beam dipole trap providing confinement in the radial axis. A lattice pulse is applied with a selected lattice depth $U_0$ for a specific exposure time $t$. The detection phase is comprised of a counter-propagating Raman velocity-selective $\pi$-pulse given at a specific two-photon detuning, selecting a narrow velocity class and transferring the atoms contained in it to the upper ground-state level~\cite{Moler1992} (fig.~\ref{Fig:Gaussians}). These atoms are then imaged using state-selective absorption imaging.

\begin{figure}
\centering
	\includegraphics[width=1\textwidth]{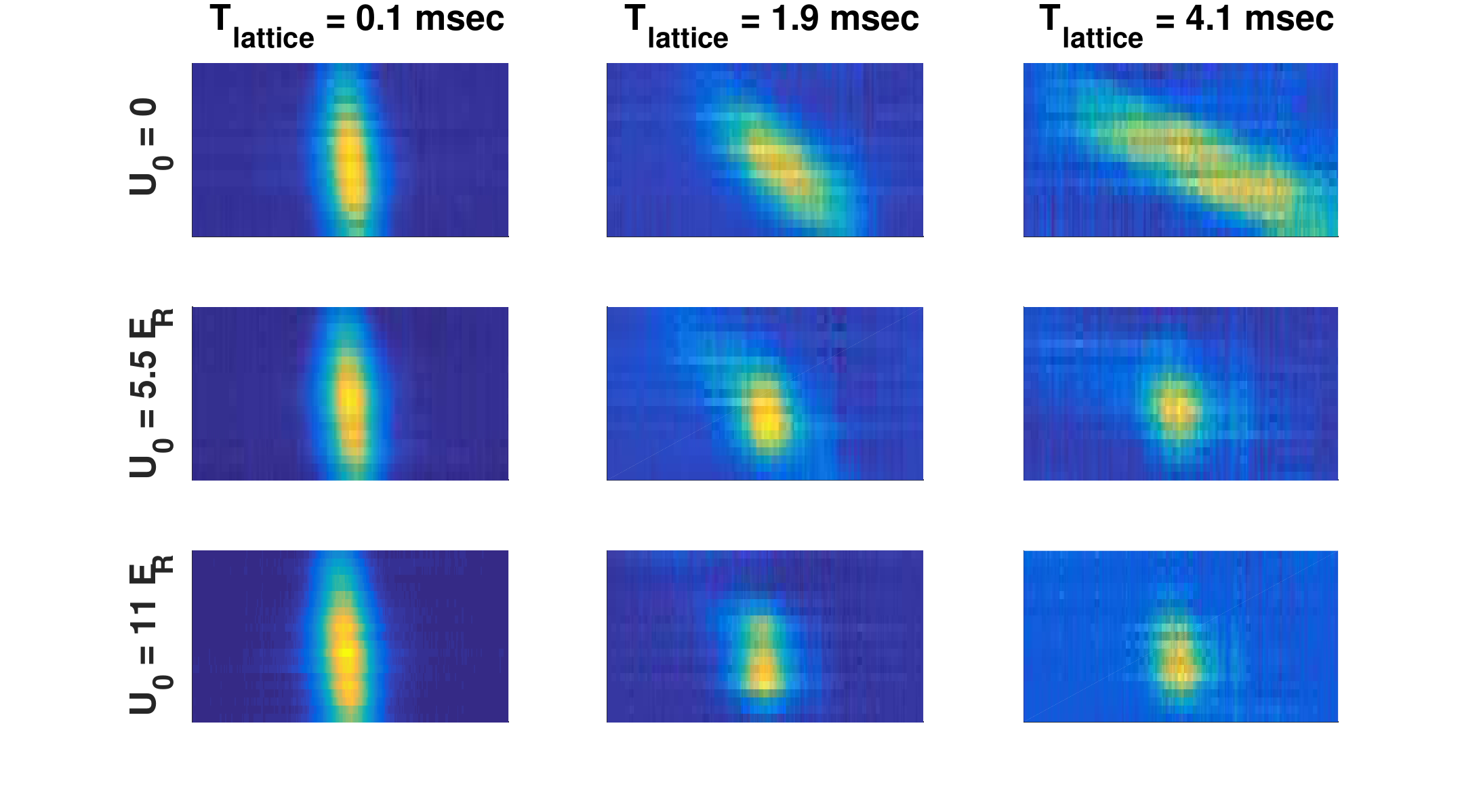}
	\caption[Selected raw data images]{Selected raw data images, for three lattice powers (rows) and three exposure times (columns). Position is on the vertical axis and velocity on the horizontal axis. The buildup of correlations in the ballistic ($U_0 = 0$) regime is clearly visible, as is the suppression of correlations in the regime where $U_0 > 0$.}
	\label{fig:some_raw_data}
\end{figure}

An example of the phase-space images obtained can be seen in figure~\ref{fig:some_raw_data}. The shearing of the phase-space in the ballistic case ($U_0=0$) is clearly visible, as is the fact that correlations are suppressed for stronger lattices.

\subsection*{Extraction of the correlation}

To quantitatively analyze the position-velocity correlations we use the definition
\eq[eq:correlation_definition]{C_{xv}(t)\equiv\frac{\braket{\delta\vec{x}(t)\cdot\delta\vec{v}(t)}}{\sqrt{\braket{\delta\vec{x}^2(t)}\braket{\delta\vec{v}^2(t)}}},}
where $\braket{...}$ denotes the ensemble average and $\delta\vec{x} = \vec{x}-\braket{\vec{x}}$. The normalization bounds the correlation between 1 and -1. If the PSD distribution is known, one can replace the ensemble average with a sum over the PSD. We extract the information about the correlations in two different ways:

\begin{enumerate}
	\item direct calculation from the raw data, thresholded by 10\% of the maximal pixel (all pixels with values $< 10\%$ of the maximal pixel are set to zero)
	\item extracting the asymmetry parameter from the data by finding the
center using fits on the one-axis-integrated data and dividing into quadrants. 
\end{enumerate}

\begin{figure}
	\centering
		\begin{subfigure}{.5\textwidth}
		  \centering
		\includegraphics[width = \textwidth]{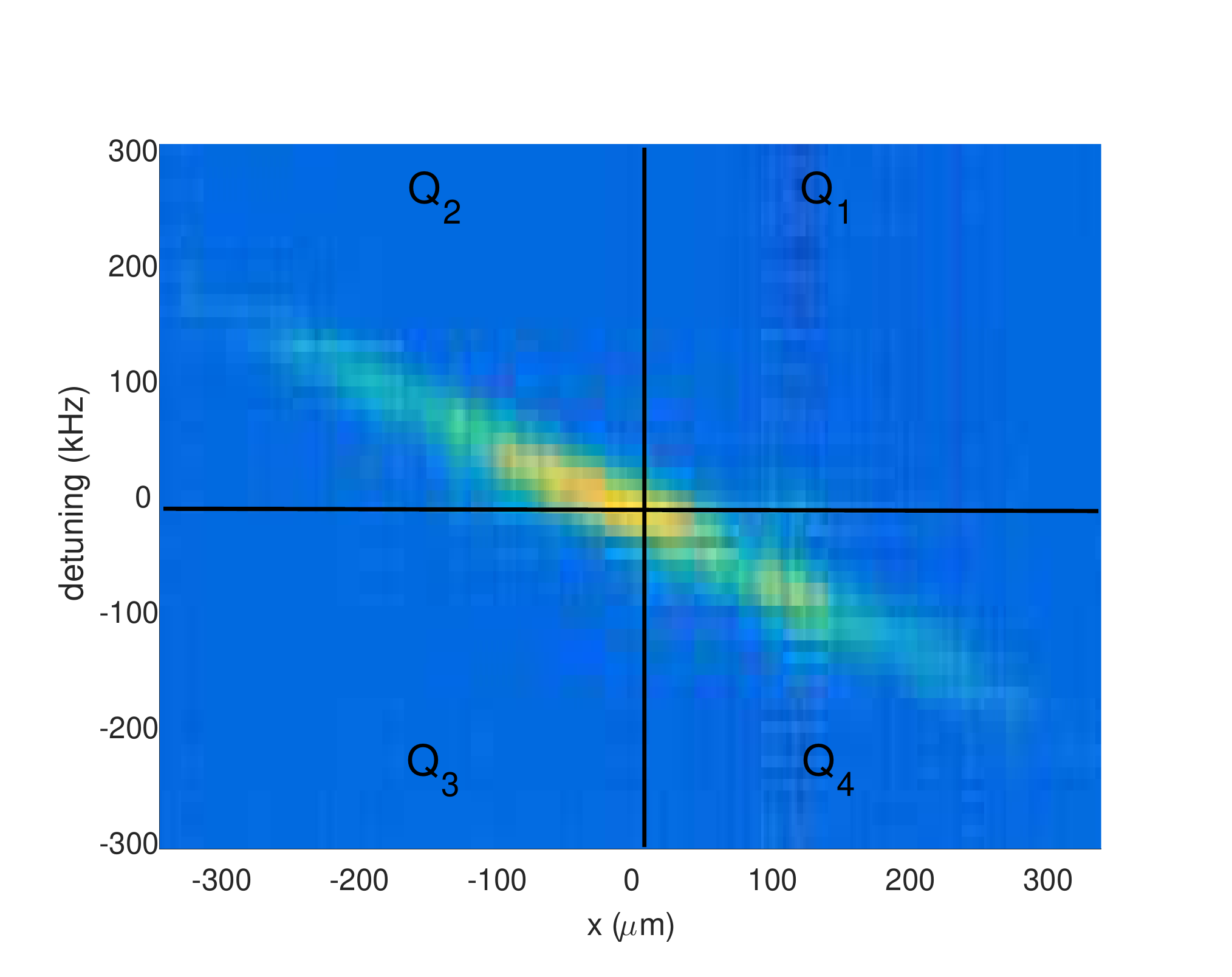}
		\end{subfigure}%
		\begin{subfigure}{.5\textwidth}
		  \centering
		  \includegraphics[width = \textwidth]{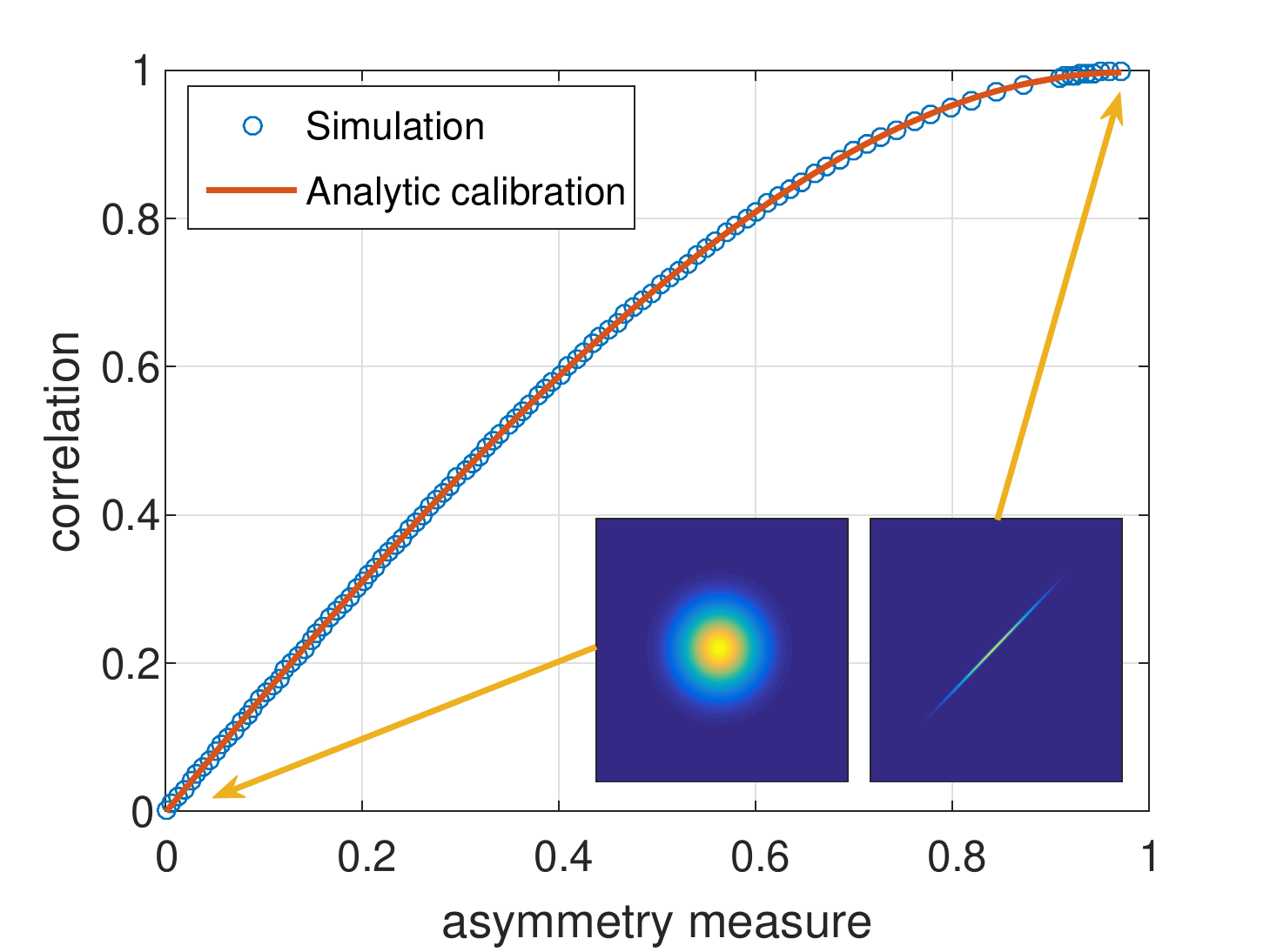}
		\end{subfigure}
	\caption[High resolution phase-space image with division into quadrants]{\textbf{(left)} High resolution (low Rabi frequency, 30 averages per slice) phase-space image with division into quadrants. Here the vertical axis is the two-photon detuning of the Raman beams. The asymmetry measure is defined as $AS = \frac{Q_2+Q_4-Q_1-Q_3}{Q_{\text{tot}}}$. \textbf{(right)} The calibration curve of the correlation as a function of the asymmetry measure. Insets show the simulated PSDD for $C=0$ and $C=99.9\%$.}
	\label{fig:High_res_long_time_30_averages}
\end{figure}

The asymmetry method works by extracting an asymmetry parameter from the data by finding the center using fits on the one-axis-integrated data, dividing into quadrants, summing the pixels in each of the quadrants to obtain $Q_i$ and extracting $AS = \frac{Q_2+Q_4-Q_1-Q_3}{Q_{\text{tot}}}$. It is then transformed into correlation using an analytic calibration, in which a bivariate normal distribution with a given correlation is integrated from $-\infty$ to zero and from zero to $\infty$ to generate the $Q_i$'s, and obtain:

\eq{AS = -1+\frac{8\arccos(C)}{5\pi-2\arcsin(C)-2\arctan(\sqrt{1/C^2-1})}}

which is then inverted numerically and used for the transformation between $AS$ and $C$. An example of the division into quadrants is shown in fig.~\ref{fig:High_res_long_time_30_averages}, along with the calibration curve.

\begin{figure}
\centering
	\includegraphics[width=\textwidth]{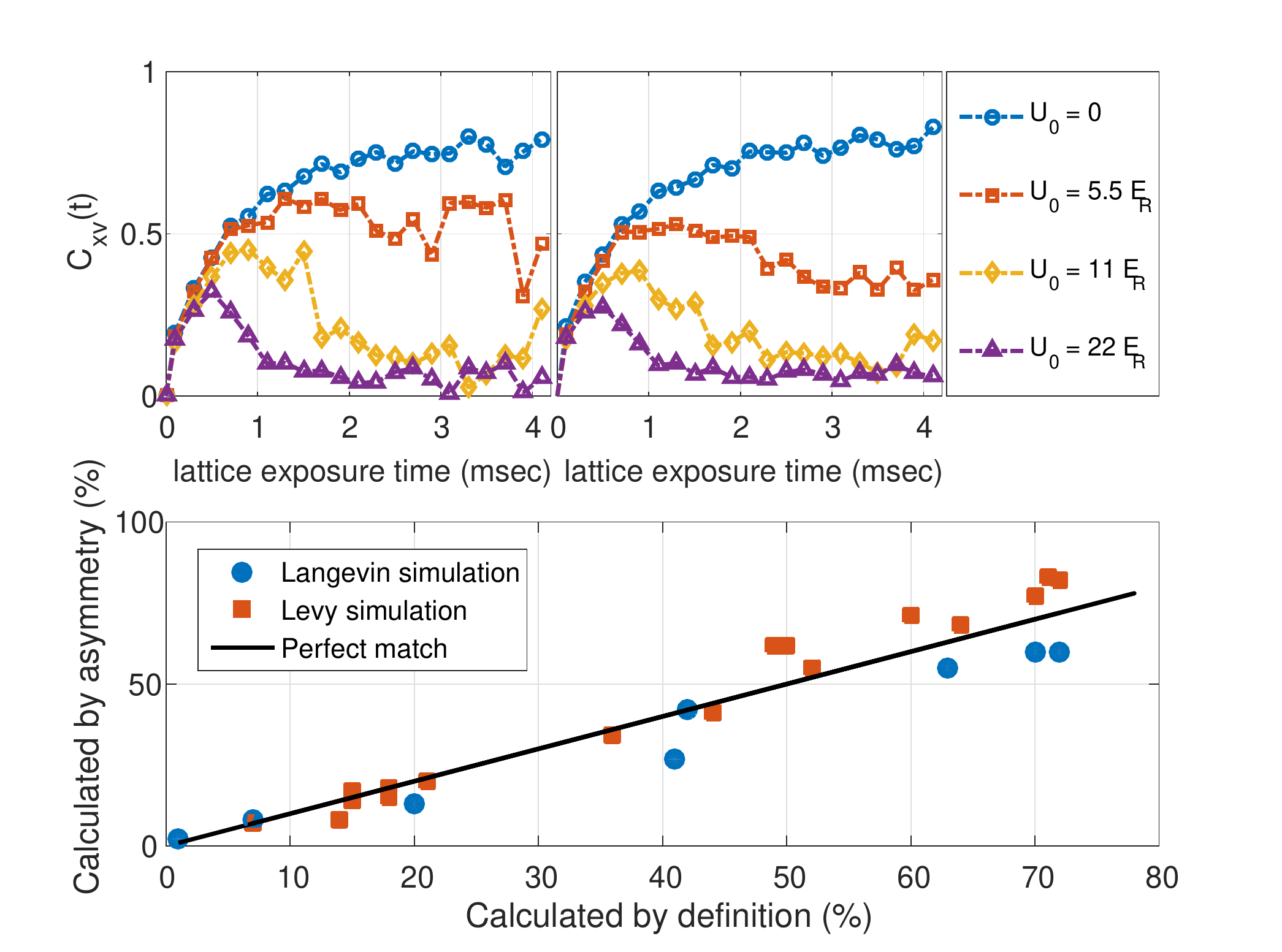}
	\caption[Position-velocity correlations]{Position-velocity correlations, directly calculated using the definition~\ref{eq:correlation_definition} \textbf{(top left)} and by using the asymmetry method \textbf{(top right)}. Here the threshold is 10\% (meaning pixels that have an intensity $< 10\%$ of the maximal pixel are set to zero). Comparing the correlations calculated using both methods on phase spaces obtained by the simulation \textbf{(bottom)}, we obtain good agreement,  especially considering that the phase spaces in question are highly non-Gaussian.}
	\label{fig:asymmetry_calibration_simulation}
\end{figure}

The correlations extracted using these methods are shown in fig.~\ref{fig:asymmetry_calibration_simulation} (top panels). One can see that the two sets behave similarly but the asymmetry one suffers less from noise. For further verification we compare the correlations extracted using the two methods on phase-spaces obtained from the two different types of simulations described in the main text. In the bottom panel of fig.~\ref{fig:asymmetry_calibration_simulation} the real correlation is plotted against that obtained from the asymmetry method. The agreement is good, especially considering that the phase spaces in question are highly non-Gaussian.

\begin{figure}
\centering
	\includegraphics[width=\textwidth]{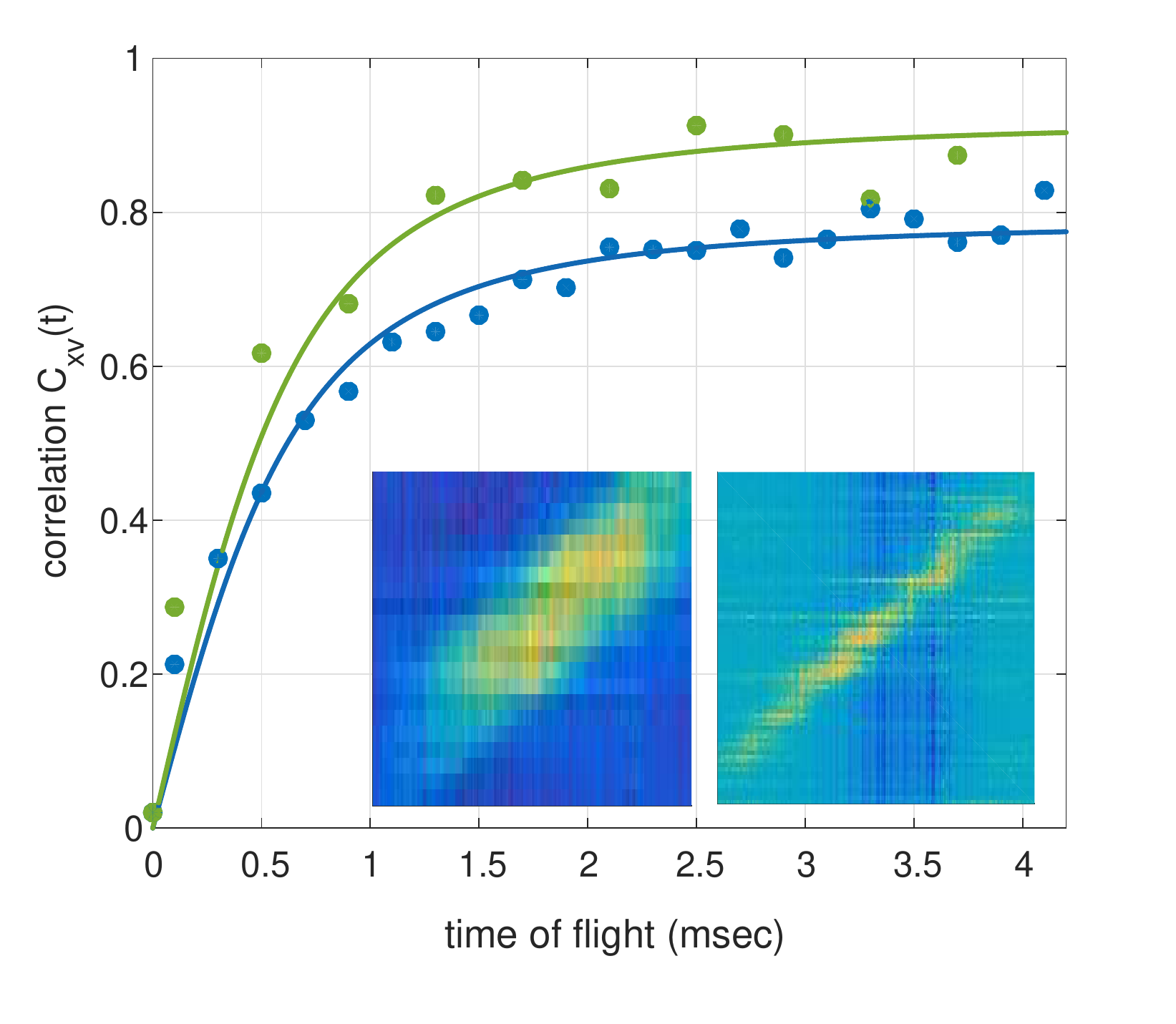}
	\caption[Varying the two-photon Rabi frequency]{Position-velocity correlations in ballistic expansion. The solid line is a fit to the analytic expression given in the paper, with the additional fitting parameter $C_\infty$ needed to account for broadening due to the finite two-photon Rabi frequency. As the Rabi frequency is reduced the broadening is less apparent, in the expense of SNR. The left (right) phase-space reconstruction corresponds to the light blue (green) data.}
	\label{fig:Broadening}
\end{figure}

\subsection*{Broadening}

a broadening effect arising from a finite two-photon Rabi frequency required to transfer enough atoms per velocity class in order to obtain reasonable SNR. The broadened correlation is a function of the ratio between the width of the Raman $\pi$-pulse~\footnote{assumed to be Gaussian for the sake of calculation simplicity} (\ie the Rabi frequency) and the standard deviation of the velocity distribution: $C_{xv}/C_{xv}^0 = 1/\sqrt{1+(\sigma_\Omega/\sigma_v)^2}$, where $C_{xv}$ is the broadened (measured) correlation, $C_{xv}^0$ is the original correlation, $\sigma_\Omega$ is the width of the Rabi $\pi$-pulse in units of velocity and $\sigma_v$ is the width of the velocity distribution. This allows for a rescaling of the correlation data to account for the Rabi broadening such that the ballistic correlations saturate at unity~\ref{fig:Broadening}. The calculation is performed by convolving a bivariate normal distribution $f(x,v,C)$ with a "Rabi" pulse $\frac{1}{\sqrt{2\pi\sigma_\Omega^2}}\exp(-v^2/2\sigma_\Omega^2)$, and then calculating the "measured" correlation using the definition given in eq.~\ref{eq:correlation_definition}.

\subsection*{Fitted parameters for the interpolation function}

\begin{figure}
\centering
	\includegraphics[width=\textwidth]{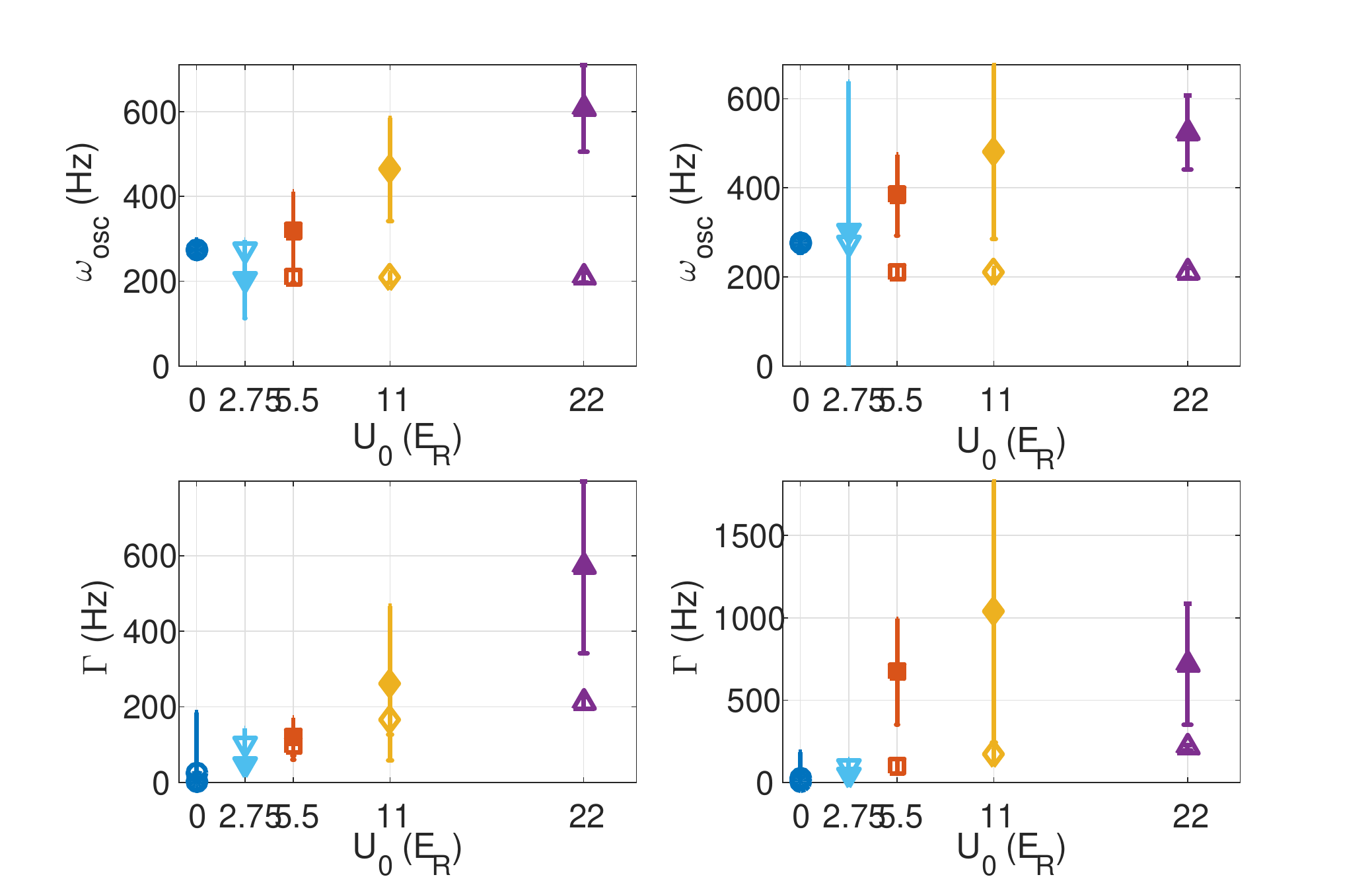}
	\caption[Fitted parameters from figure 2 of the main text]{fitted parameters $\omega_{\text{osc}}$ \textbf{(top)} and $\Gamma$ \textbf{(bottom)}, shown in full symbols, along with their errors for both the unexcluded fit of fig. 2 (a) \textbf{(left)} and the excluded fit of fig. 2(b) \textbf{(right)} of the main text. The parameters are compared to those obtained independently (empty symbols) by either a trap oscillations experiment or an exponential fit to the velocity dynamics.}
	\label{fig:Fitted_parameters}
\end{figure}

For the fit to the data depicted in figure 2 of the main text, we use the generalized interpolation function given in equation 9 of the main text. Figure~\ref{fig:Fitted_parameters} shows the fitted parameters $\omega_{\text{osc}}$ and $\Gamma$, shown in full symbols, along with their errors for both the unexcluded fit of fig. 2 (a) and the excluded fit of fig. 2(b) of the main text. The parameters are compared to those obtained independently (empty symbols) by either a trap oscillations experiment (giving a small kick to the trapped atoms, and imaging the oscillations in the trap) in the case of $\omega_{\text{osc}}$ or an exponential fit to the velocity dynamics in the case of $\Gamma$.

\subsection*{Full derivation of the normal-diffusion limit}

We calculate the position-velocity cross-correlation coefficient according to~\ref{eq:correlation_definition} for the case of simple Brownian motion. Beginning with the Langevin equation describing the instantaneous acceleration of a particle in a medium~\cite{Pathria1996,Gillespie2012},
\eq[eq:Langevin]{\dot{\vec{v}} = -\Gamma\vec{v} + \vec{\mathcal{A}}}
where $\vec{v}$ is the velocity vector, $\Gamma$ is the drag coefficient setting the timescale for transition between the ballistic and diffusive regimes and $\vec{\mathcal{A}}$ is the Langevin random  acceleration. For the sake of simplicity of notations we assume from here on $\braket{x}=\braket{v}=0$, hence $\delta x = x$ and $\delta v = v$. The numerator of eq.~\ref{eq:correlation_definition} can be calculated by noticing that-
\begin{equation}
\label{eq:diff}
	\begin{aligned}
		\der{}{t}\braket{\vec{x}\cdot\vec{v}} &= \braket{\der{}{t}(\vec{x}\cdot\vec{v})} = \braket{\vec{x}\cdot\der{\vec{v}}{t} + \vec{v}\cdot\der{\vec{x}}{t}}\\
		&=\braket{\vec{x}\cdot(-\Gamma\vec{v}+\vec{\mathcal{A}})+\vec{v}^2}\\
		&=-\Gamma\braket{\vec{x}\cdot\vec{v}}+\braket{\vec{v}^2}
	\end{aligned}
\end{equation}
Between the 2nd and 3rd line we use the fact that due to the randomness of the Langevin acceleration $\braket{\vec{x}\cdot\vec{\mathcal{A}}}=0$. The time dependence of $\braket{\vec{v}^2}$ is given by-
\eq[eq:vsq]{\braket{\vec{v}^2} = v_0^2 + (v_{eq}^2-v_0^2)(1-e^{-2\Gamma t})}
with $v_0^2 = \sigma_v^2(t=0)$, the initial variance of the velocity distribution, and $v_{eq}^2 = \sigma_v^2(t\to\infty) = \frac{3k_BT}{m}$ according to equipartition in 3D ($k_B$ is the Boltzmann constant, $T$ is the temperature and $m$ the mass of the particle). Substituting into eq.~\ref{eq:diff} and solving under an uncorrelated initial condition $\braket{\vec{x}(t)\cdot\vec{v}(t)}\vert_{t=0} = 0$ we get:
\eq[eq:xv]{\braket{\vec{x(t)}\cdot\vec{v(t)}} = \frac{e^{-2\Gamma t} \left(e^{\Gamma t}-1\right)\left(v_{eq}^2\left(e^{\Gamma t}-1\right)+v_0^2\right)}{\Gamma}}
Next we require the terms in the denominator of~\ref{eq:correlation_definition}. The velocity was already presented in~\ref{eq:vsq} and we just need to take its square root. The position, however, requires the solution of the differential equation-
\eq{\dsq{\braket{\vec{x}^2}}{t}+\Gamma\der{\braket{\vec{x}^2}}{t} = 2v_0^2e^{-2\Gamma t} + 2v_{eq}^2(1-e^{-2\Gamma t})}
under the initial conditions $\braket{\vec{x}^2(0)} = x_0^2$ and $\der{\braket{\vec{x}^2}}{t}\vert_{t=0}=0$. This gives:
\eq[eq:xsq]{\braket{\vec{x}^2} = \frac{e^{-2\Gamma t}\left(v_0^2 \left(e^{\Gamma t}-1\right)^2+v_{eq}^2\left(e^{2\Gamma t}(2\Gamma t-3)+4e^{\Gamma t}-1\right)+\Gamma^2x_0^2e^{2\Gamma t}\right)}{\Gamma^2}}
Now we switch to the following unitless parameters:

\begin{itemize}
	\item $\eta\equiv\frac{v_0^2}{v_{eq}^2}$ denoting the deviation of the initial velocity distribution width from its equilibrium value (set to unity in the main text for simplicity).
	\item $\omega_{osc}\equiv \sigma_v(0)/\sigma_x(0)$, the previously defined trap oscillation frequency that sets, under thermal equilibrium, the ratio of the initial conditions of the velocity and the position.
\end{itemize}

and combine everything to get the main result:

\begin{equation}
\frac{\left(e^{\Gamma  t}-1\right) \left(\eta +e^{\Gamma  t}-1\right)}{\sqrt{\left(\eta +e^{2 \Gamma  t}-1\right) \left[(\eta -1)+e^{2 \Gamma  t} \left(\eta  \left(\frac{\Gamma }{\omega }\right)^2+(\eta +2 \Gamma  t-3)\right)-2 (\eta -2) e^{\Gamma  t}\right]}}
\end{equation}

Setting $\eta=1$ we reobtain eq. 8 of the paper:

\begin{equation}
\label{eq:main}
	C_{xv}(t) = \frac{e^{-\Gamma t/2}\left(e^{\Gamma t}-1\right)}{\left[2+e^{\Gamma t}\left((\Gamma/\omega_{\text{osc}})^2+2\Gamma t-2\right)\right]^{1/2}}
\end{equation}

\begin{figure}
\centering
	\includegraphics[width=\textwidth]{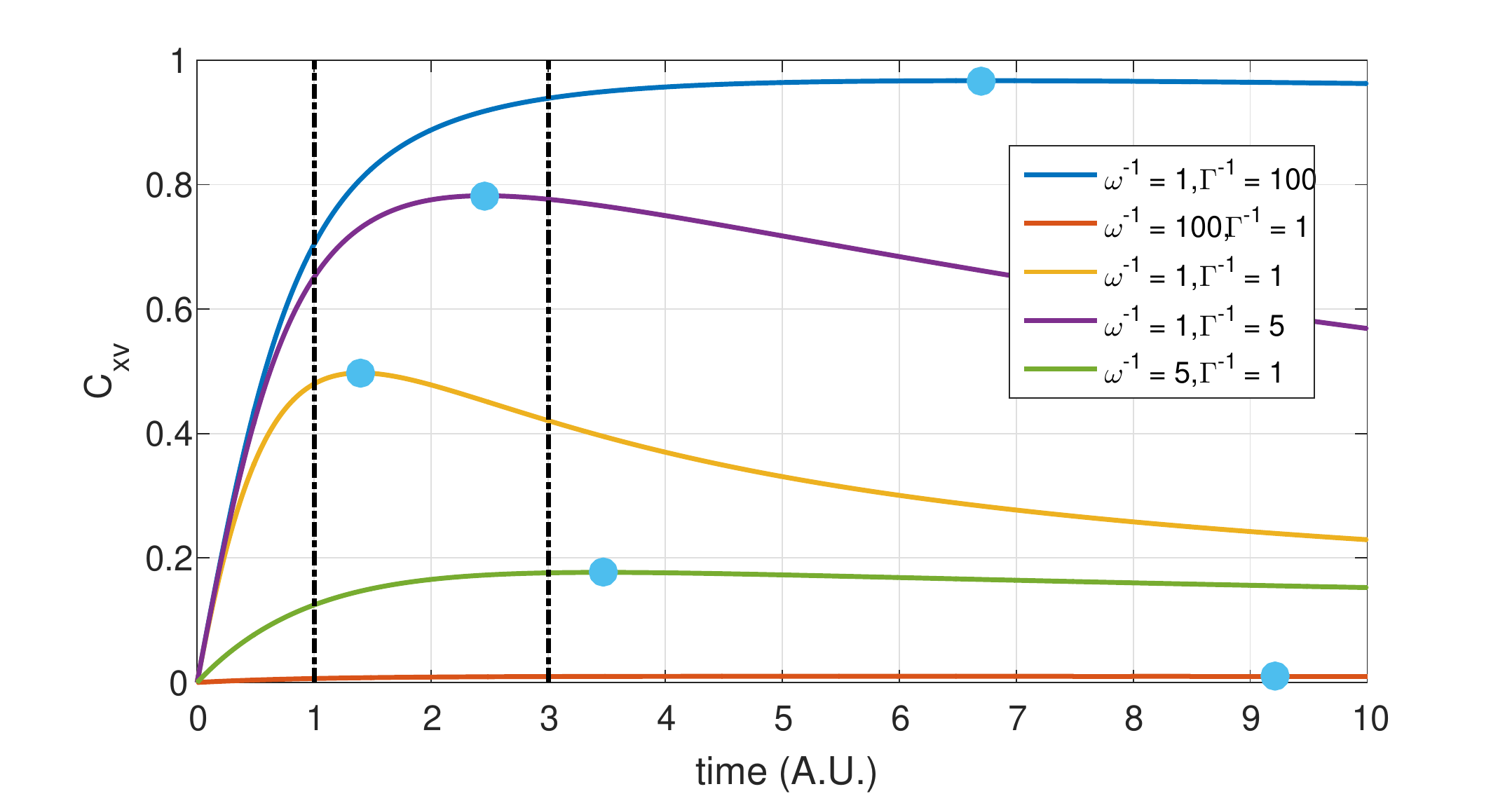}
	\caption[Plot of eq.~\ref{eq:main}]{Analytic plot of eq.~\ref{eq:main} as a function of time for different values of $\omega_{\text{osc}}$ and $\Gamma$. The light blue circles indicate the maximum calculated by eq.~\ref{eq:max_time} and the dashed black lines are the approximation obtained by taking the average of the two timescales $\left(1/\omega_{\text{osc}}+1/\Gamma\right)/2$.}
	\label{fig:timescales}
\end{figure}

Analyzing the timescales inherent to the system yields the following insight: There exist two time \emph{scales} and two temporal \emph{scalings}. The buildup scales linearly in time and saturates at unity with a timescale of $1/\omega_{\text{osc}}$. The decay scales like $t^\gamma$ with a timescale of $1/\Gamma$. The transition between the buildup and decay occurs at a timescale $\tau_m$, which is approximately the average $\left(1/\omega_{\text{osc}}+1/\Gamma\right)/2$. It is henceforth evident that observing the short-time correlation dynamics requires $\tau_m$ to be within the measurement time. To be more precise, maximization of eq.~\ref{eq:main} yields the following transcendental equation for $\tau_m$, the maximal correlation time:

\begin{equation}
\label{eq:max_time}
	\Gamma\left(\Gamma+2\tau_m\omega_{\text{osc}}^2\right)-2\omega_{\text{osc}}^2\sinh(\Gamma \tau_m)=0
\end{equation}

that can be solved numerically to give the exact time and value of the maximal correlation in this model. In fig.~\ref{fig:timescales}, we plot eq.~\ref{eq:main} as a function of time for different values of $\omega_{\text{osc}}$ and $\Gamma$. The light blue circles indicate the maximum calculated by eq.~\ref{eq:max_time} and the dashed black lines are the approximation obtained by taking the average of the two timescales $\left(1/\omega_{\text{osc}}+1/\Gamma\right)/2$, showing that it is a valid approximation. These insights hold also for the case of anomalous diffusion, where the only difference with respect to this analysis is the scaling of $t^\gamma$ instead of the specific $t^{-1/2}$.

\pagebreak
\bibliographystyle{apsrev}
\bibliography{correlations_supplementary}